\documentclass{acmart}

\AtBeginDocument{%
  \providecommand\BibTeX{{%
    \normalfont B\kern-0.5em{\scshape i\kern-0.25em b}\kern-0.8em\TeX}}}

\setcopyright{acmcopyright}
\copyrightyear{2021}
\acmYear{2021}

\acmConference[CTAM21]{CHI'21 Workshop on Technologies to Support Critical Thinking in an Age of Misinformation}{May 09, 2021}{Melbourne, AU}

\definecolor{green}{rgb}{0.2,0.5,0.2}
\definecolor{orange}{rgb}{1,0.5,0}
\definecolor{lightgray}{rgb}{0.4,0.4,0.4}

\usepackage[acronym]{glossaries}
\newacronym{wysiwyg}{WYSIWYG}{What You See Is What You Get}
\newacronym{ui}{UI}{User Interface}
\newacronym{ux}{UX}{User Experience}

\begin{document}



\title[Piercing the Veil]{Piercing the Veil: Designs to Support Information Literacy on Social Platforms}

\author{Jan Wolff}
\email{jan.wolff@hci.uni-hannover.de}
\affiliation{%
  \institution{Human-Computer Interaction Group, Leibniz University Hannover}
  \streetaddress{Appelstraße 9A}
  \city{Hannover}
  \state{Lower Saxony}
  \country{Germany}
  \postcode{30167}
}


\author{}

\begin{abstract}


In this position paper we approach problems concerning critical digital and information literacy
with ideas to provide more digestible explanations of abstract concepts
through interface design. In particular, we focus on social media platforms where
we see the possibility of counteracting the spread of misinformation by providing
users with more proficiency through our approaches.
We argue that the omnipresent trend to abstract away and hide
information from users via UI/UX design opposes their ability to self-learn.
This leads us to propose a different framework in which we unify elegant and simple
interfaces with nudges that promote a look behind the curtain. Such designs
serve to foster a deeper understanding of employed technologies and aim to
increase the critical assessment of content encountered on social platforms.
Furthermore, we consider users with an intermediary skill level to be
largely ignored in current approaches, as they are given no tools to broaden their knowledge without
consultation of expert material. The resulting stagnation is exemplified
by the tactics of misinformation
campaigns, which exploit the ensuing lack of information literacy and critical thinking.
We propose an approach to design that
sufficiently emancipates users in both aspects by promoting a look behind
the abstraction of UI/UX so that an autonomous learning process is given the chance
to occur. Furthermore, we name ideas for future research within this area that
take our considerations into account.


\end{abstract}



\maketitle

\section{Introduction}

The surge of misinformation campaigns and conspiracy theories during the
COVID-19 pandemic are, in part, symptomatic of a severe lack of critical
digital literacy in parts of the online population. We want to propose the
argument that this represents just a very visible tip of the iceberg. Ever
since home computing became ubiquitous, the main target of interface design has
been to initially make computers, then the internet,
and by proxy a vast amount of information easily digestible.
Yet, as became apparent in the debate on the term \textit{digital native}, mere
exposure to digital systems does not imply a deep understanding of the
technology at hand \cite{2008digitalnativedebate}. Nevertheless, especially
since the widespread introduction of smartphones and social media, computer
usage and interconnectedness is on a stable path to become globally
ubiquitous. In 2018 for the first time more than half of the global population
were using the internet in a continuously rising trend
\cite{pub_series-dataset-64cb0e71-en}. But this development is not
accompanied by a simultaneous in-depth education on digital systems. As a
matter of fact, it is mainly driven by the introduction of systems that keep
their internals hidden. In part to explicitly avoid a critical examination.

One visible manifestation of the problems posed by this fact is found in online
spaces where users are provided with a seemingly endless influx of information
that is highly personalized~\cite{10.1145/2959100.2959190}.  Within the context
of the \textit{attention economy}~\cite{attentioneconomy} this serves to
increase the time spent lingering on each platform in order to maximize exposure
to advertisements. A negative side-effect
arises due to the relative ease with which content can be shared and one's own
material can be circulated. The lack of information about each platforms' inner
workings therefore runs the risk of leading users to draw wrong conclusions
about why content recommendation systems expose them to certain media
pieces~\cite{algorithmicimaginary}. Unknowingly, users are subjected to highly
sensationalized content that leads to longer interaction with the platform.
The ease of sharing, discussing and creating media pieces encourages one's own
tendency to sensationalize, as such content is rewarded by being recommended to
a larger audience. There are indications that these mechanisms play a substantial
role when it comes to problems with radicalization on online platforms
through reinforcement of fringe opinions within communities~\cite{nytinsurrection}. The personalization
of presented content further causes the formation of \textit{filter bubbles}~\cite{Pariser2011TheFB},
which present an overly skewed view of the global
community, leading to incorrect assumptions about the prevalence and
shared agreement of expressed opinions.

\subsection{Problem Statement}
\label{sec:problem_statement}


With these previous assumptions in mind, we see the problem of lacking critical
media literacy to be twofold. For one, rewarding sensationalized news and
opinion pieces disincentives unopinionated factual statements, which then leads
to a lack of fact checking as it solely reinforces and strengthens previously
held viewpoints. But secondly, users are given nearly no agency to assess and
explore different opinions once they are part of a filter bubble. One of the
few places where exposure to other segregated communities might occur is in
globally trending topics, where---once again---sensationalized content is
rewarded by being exposed to a larger set of people. Such circumstances are
hardly beneficial for healthy and factual discussion, because the most
visible opinions are strongly influenced by the prevalent sentiments found in
their respective communities and express often incompatible
viewpoints. This serves to create environments in which factual evidence is
given lesser weight than non-factual content, as it is surrounded by other
highly charged statements if it is expressed by another social group.


This area is often neglected when
it comes to solving the problem of fake news. We see a chance in
giving users more agency about exploring opinions and giving more visibility to the
factors that contribute to their selection of content. As a
first step, we address the increasing opaqueness of user interfaces found on
social media platforms. For this reason we explore examples of designs
that---while user-friendly---prevent any engagement with the layers below
them.

\section{Addressing Social Platforms}
\label{sec:social}




Certainly the design of commercial social platforms is based most
strongly on economic incentives. Some of the problems thus stem from
\textit{recommender systems}, which present users with content
that is vast in scale, high in velocity\footnote{As one of the five \textit{V}'s
of big data, velocity refers to the rate at which new content is created.} and automatically
selected to cause long periods of interaction with the platform.
Visible indications explaining the reasoning behind
recommendations are often missing, which is
especially concerning given that a main factor for pre-selecting
content is the user's predicted interaction with it. Thus, unbeknownst to
users, they are presented with controversial information in order to elicit
interaction with the content \cite{bucher2017affordances}. As this property is not
communicated by designs, users who are missing this
insight have been found to assume
that they are simply presented a balanced selection of the content they are
subscribed to \cite{10.1145/3173574.3173677}.

Similarly, grasping the scale of a platform's user base is not always possible
within the bounds of the platform itself. Taking \textit{Twitter} as
an example, only few tools are provided that allow gaining an awareness of how
representative one's selection of connected profiles is in relation to the
platform's general population. The only methods available at the time of
writing are a global search, the list of trending topics, and an indicator on
each profile page showing the list of mutual followers. However, getting any
meaningful large-scale information using only these tools is impossible, simply
due to the required amount of manual work. Furthermore, knowing about the existence of
certain peer groups is a prerequisite for searching them, inverting the
directionality that should be provided.

When looking beyond social media, other user interfaces show similar
tendencies when analysed. For example, an ongoing trend in desktop interfaces
is to be increasingly less verbose about processes that are performed. This
poses the risk of leading users to take several aspects presented by the
GUI as uncontrollable. Thus, while simplifying the initial process
of learning how to interact with such systems, any mental models and
experiences are strictly coerced into the set of tasks provided by the UI. No
elements indicate that deeper knowledge could be acquired, possibly engendering
a mental state similar to the Dunning-Kruger effect
\cite{mahmood2016dunningkruger, GIBBS2017589}, as it is impossible to perform
an offhand assessment of one's own expertise. This bears the risk of
contributing to factors that make users more susceptible to misinformation
campaigns, especially concerning the content encountered on social platforms. Yet,
other aspects of digital systems can be affected as well. For example, assumptions made
about privacy, security, or malware.

\subsection{Related Work}

Previous works exist in which methods for increasing awareness towards
the underlying data, algorithms, and applications have been explored.
Common methodologies either give more context to the content found on
different platforms or directly support a deeper understanding
of the technological and societal mechanisms at play.

\subsubsection{Opinion Space}

This work by Faridani et al.~\cite{10.1145/1753326.1753502} addresses the
problem of assessing the prevalence and relationship of opinions. In their
design, answers to a primary question are visually laid out in a
two-dimensional space based on participants' responses in a supplementary
5-question opinion profile. Through a principal component analysis the
5-dimensional answer space is projected onto a two-dimensional plane.
Additionally, comments with a high rating stand out visually, guiding users
towards them, while an overview of the diversity in opinions is maintained. In a
user study, the interface has been found to increase dwell times in comment
sections and to cause greater respect and agreement concerning encountered opinions
when compared to a classic, time-sorted list interface. This shows the benefits
of giving more agency and tools to users when dealing with discussion platforms.

\subsubsection{Balancer Browser Extension}

The \textit{Balancer} extension~\cite{Munson_Lee_Resnick_2013} provides a nudge
with the intent to diversify the selection of news sources consulted by the
user. A prominently positioned browser widget displays a stick figure carrying
a block in each hand, with their sizes representing the amount of either
conservative or liberal media pages visited. An uneven weight
distribution between the two boxes causes the figure to tilt to the corresponding side.
In a supplementary study a ``small but real''~\cite[p.~427]{Munson_Lee_Resnick_2013}
effect could be seen in the
behaviour of participants with the visualization enabled. The work
serves as an example of a method that gives unopinionated feedback concerning the user's
online behaviour and presents a design space that future approaches
can take into consideration.

\subsubsection{Explanations for Supporting Algorithmic Transparency}

In \cite{10.1145/3173574.3173677} Rader et al. explore methods of providing
explicit explanations for the algorithms in use by a given system. A
representative imaginary student admission algorithm was used to evaluate the
effectiveness of different approaches. Depending on the experiment conditions,
either a detailed visualization of the mapping between input and output
(\textit{white-box}) or the algorithm's binary result (\textit{black-box}) was
displayed. Additionally, participants could either \textit{interactively}
adjust the input parameters to learn their contribution to the result
or were presented with only \textit{static} values.
A user study showed that, while participants did not self-report a heightened
understanding of the system, the increased verbosity did have a positive impact
when objectively measured.
This fact may stem from the aforementioned Dunning-Kruger effect. People are only
able to truly reflect on their knowledge once confronted with the system's internals.
Thus, even if they objectively gain knowledge, their self-assessment does not change
as they are made aware of further aspects that they do not know about.
Such results can serve as a pointer for future research when considering
similar explanatory methods.

\subsubsection{Privacy Nudges}

Almuhimedi et al. explore an idea in which they
explicitly nudge users to review their app permissions by telling them, for
example, how often their location has been requested by an app~\cite{10.1145/2702123.2702210}. The
supplementary permission manager \textit{AppOps} further streamlines the
process of permission configuration. A study
could show that such nudges, in tandem with the additional management interface,
cause significantly more users to review and impose
limits on the data that is accessible to applications.
The basic idea of these nudges can therefore serve as
a basis for communicating even more detailed information to users.

\section{Emancipatory Design}

We now seek to propose ways in which the issues mentioned in \autoref{sec:problem_statement} can be
counteracted. In what we call \textit{emancipatory design} we envision
interfaces that, while still providing a low barrier of entry, give more insight
into the systems they interact with. Such
interfaces can, of course, not display the entirety of each system but only a
carefully chosen subset of aspects beyond their level of abstraction.  This
way, users are made aware of the fact that they could benefit from actively
searching out more information about the technology at hand. Going
further, such interface designs may directly include possible methods of
knowledge acquisition. They may also allow autonomously exploring aspects of the underlying
system. Designs necessarily need to coincide with
the concepts of minimalism and usability in order to present a step beyond
the current state-of-the-art.

Preferably, this approach is to be holistic. While individual systems are likely to
benefit from users gaining a more positive disposition towards actively
learning more about them, social implications could go far beyond this. At
best, users are at least partially aware of all aspects concerning their
everyday systems. Mental models then paint a simplified picture that
fits reality closely enough to prevent situations in which uneducated guesses
have to be made. We expect that, once one veil is lifted, users begin to
assess other aspects more critically as well. Such developments could then help
to counteract over- or underconfidence and protect users
against false advertising, fake news, and other types of misinformation. We
believe that the greatest benefit is to be had if such concepts are adopted in a consistent way in many
different types of interfaces that users come into contact with.

We assume that several approaches need to be considered. While some
interfaces provide opportunities to implement additional interactions used for
introducing low-level concepts, others may not. Such cases can arise due to
concepts that do not fit well into digestible visualizations or because of
platforms that explicitly attempt to hide design choices. Careful analysis has
to be conducted when considering the latter, as such approaches will
necessarily obstruct the user's workflow within the application itself.  Hence,
if a positive effect is to be had, the delivery must be kept short and
unobtrusive.

\subsection{Examples}

We will now give some examples for designs that could serve as a basis for
future research. First, we present interactions that can be introduced
into existing interfaces, followed by designs that are external to the system
they aim to explain.



\subsubsection{Exploring Filter Bubbles}

As previously explained, one key factor of increased exposure to misinformation are
\textit{filter bubbles} \cite{Pariser2011TheFB}, which cause an apparent amplification of
otherwise fringe opinions. We therefore consider designs in which the
existence of such bubbles and the partisanship within is given increased
visibility.
To avoid possible negative implications, care should be taken to perform visualizations
as objectively as possible, because they should only serve as a tool
that allows users to gain an understanding of the otherwise difficult to
comprehend network graph between profiles.  Furthermore, selective hiding or
ranking of such bubbles should not occur, as existing research suggests
exposure to opposing viewpoints to be an aide in fostering accurate beliefs
\cite{10.1037/a0015701}.

One approach to study in future research could combine the design of
\textit{Opinion Space} with an automated positioning of profiles via the social
graph. Faridani et al. suggested the need for an easy to understand scale in
the visualization, for which they proposed displaying selected public
personalities as landmarks~\cite{10.1145/1753326.1753502}. When considering the
application of this idea on \textit{Twitter}, this design can easily be
translated by selecting well-known profiles and always displaying those as
landmarks in the graph.  Such visualizations may be enabled manually through an
action presented as \textit{zooming out}. Users are first presented their
regular, self-curated feed, from which they can zoom out to see
a slightly simplified overview of the whole network using selected
profiles as landmarks.  Given such a view, one could imagine being able to
effectively \textit{zoom in} and experience feeds or discussions from the
perspective of another bubble, which can allow people to have an insight into sentiments
outside their peer group. 

\subsubsection{Explaining Recommender Systems}

While the aforementioned idea plays strongly on social groups, another concept
can be applied to recommendation systems. Considering video platforms such as
\textit{YouTube}, automated content moderation is largely based on the
selection of videos the particular user has watched
in the past \cite{10.1145/2959100.2959190}. In order to
better represent the existence of clusters in such a system, users should be
given the chance to understand what selection of content contributed most to
each recommendation. In a possible solution, the list of recommendations and a
history of content viewed in the past could be combined. To facilitate quickly
assessing the information at hand, related entries in the list of past content
would be grouped. Then, for each recommended item the connections to either
individual entries or whole groups can be displayed.

While such visualizations and explanations can never paint a complete picture
and go into deep detail, a positive effect may still be had even with
some simplification. Edwards and Veale suggest that, even when given only superficial
information about machine learning practices, people are sufficiently
able to picture a ``model-of-a-model'' \cite{edwards2017slave} so
that meaningful conclusions about the system can be drawn.

\subsubsection{Auxiliary Context Aware Learning}

As already mentioned, factors can be at play that do not permit such measures
to be directly embedded in an application's interface. Hence, approaches need to be considered
in which information stems from applications external to the one that is to be
explained. These ideas must go hand in hand with context awareness, in order to
be adequately effective and not be perceived as a nuisance. Behavioural
analysis should be applied to pinpoint moments during which users are not
actively performing a task, but in which they are available to interact with a
knowledge gathering application for a short time. While research supporting
such systems exists~\cite{Dingler2015Proxemics}, an additional dimension to
support such applications could be the users' level of knowledge about the
system. In such cases, the system could monitor whether intermediary users
operate on a suboptimal plateau of performance. If they do, new interactions
or concepts could be introduced step-by-step. However, further research is
required to assess if---and to what degree---such information can be gathered automatically.



One holistic approach could make use of \textit{gamification}. Given a
\textit{Jeopardy}-style selection of topics and corresponding pieces of
information, users could be motivated to gather knowledge about a wide
assortment of topics. In fact, applications with comparable approaches already
exist. One example would be the general knowledge gathering app
\textit{getucated}\footnote{\url{https://getucated.de}}, which focuses mostly
on \textit{history}, \textit{philosophy}, and
\textit{culture}. Combined with context awareness, such applications could put
an emphasis on areas that are relevant to the specific user and in which they
display a lack of knowledge. In order to further enhance the appeal of the idea,
one could evaluate whether a personification of the knowledge gathering application
via a digital assistant has a positive impact on user retention.



\subsubsection{Judgemental interfaces}






Given the fact that information given by external sources is free to emphasize
negatives, more strongly worded language can be employed.
One could, for example, measure the time users spend clicking through video recommendations
on \textit{YouTube}. After a while, a notification could be shown explaining
that the recommended content is carefully chosen to maximise the
time spent on the platform in order to expose users to as many ads as possible.
Such messages should be shown regardless of the particular content on screen
to prevent users from assuming a bias concerning certain media pieces. Yet,
in case users become entrenched within single topics, such notifications could
serve as a nudge to take a step back from the content.

Going further, one could imagine texts that take into account even more
information from, for example, \textit{Google}'s advertising profile of users.
This information could then serve as a basis for messages such as:
``Google knows you are in a relationship, did you?''
or ``Do you visit this restaurant often? Google thinks so.''
This idea is related to the aforementioned
\textit{privacy nudges}~\cite{10.1145/2702123.2702210}.


\section{Discussion}



The above-mentioned ideas reflect a subset of possible interface designs that
take emancipatory approaches into account, with a focus specifically on social
platforms. Therefore, mainly their potential social implications are
considered, with the emancipation of users regarding technical aspects being
less of a goal in this context. The particular selection of examples represents
ideas which we deem feasible in their realization. In the first two examples we
consider designs that would---beyond a prototyping stage---require cooperation
with social platforms themselves. The other two designs allow to relax this
requirement, as their basic functionality stems from locally analysing user
behaviour and serving an assisting purpose. We thus assume those examples to
present a valid starting point, with their prototypical evaluation being
feasible for first studies and evaluations, due to their concepts being similar
in scope to existing related work.

We envision that, once design approaches include the concept of presenting a
digestible look behind the veil of systems they abstract away or actively
offer an opportunity to seek out information, novice users are given
an incentive to learn about aspects concerning the systems they use. They do
not need to become experts, they just need to gain enough knowledge to develop an
increased ability to critically assess digital systems. We see the chance that
such developments can provide users with enough agency to overcome adversarial
practices that prey on their lack of critical knowledge. For social
platforms this should, at best, manifest in an awareness for the existence of
social bubbles, the aim of automated recommendation systems and the impact
these factors have on the selection of content. All aspects combined have the
potential to counteract increasing segregation into sub-communities within
platforms and thus factors that play a significant role in
susceptibility to fake news. Going further, other areas of personal
computing can benefit from an increased awareness as well. For example, an
increased resistance of users towards data harvesting practices or premature
obsolescence of otherwise functional systems.

Notably, our ideas do not directly involve automated fact checking or elements from
\textit{inoculation theory}. Rather, we see part of the reason for the success
of online misinformation campaigns in a manifestation of \textit{learned
helplessness}~\cite{learnedhelplessness} due to increasingly concise interface
designs in which users are taught to act passively. We theorize that users
extend newly gained critical thinking skills concerning platforms and devices
towards the content presented in them.
In the same manner, we argue against automated methods of confronting
users with content sourced from outside their filter bubble. However, exposure
to only partisan sources reinstating one's own world-view and increased
polarization accompany each other~\cite{10.1111/j.1460-2466.2010.01497.x} and
fostering communication between boundaries of in-groups has been proposed as
beneficial~\cite{10.1177/0270467610380011}. In this, we see an excellent
example where increasing the visibility of such groups and providing tools for people to
broaden their scope themselves is more beneficial compared to the automated
counterpart, in which we see a hazard of users assuming malice. We see the
possibility of passively making people aware of their bubble's scope and its
true size to be a viable option in preventing social segregation in online spaces
worthy of further research, whereas relying solely on automated
techniques merely shifts the problem space. In
particular we see aspects of inoculation theory already being used to instil
beliefs in conspiracy theories, such as nurturing a distrust in reputable
sources or preemptively giving non-factual counterarguments to any statement that
would disprove the theory\footnote{Instilling distrust in mainstream media and
scientific sources is a cornerstone in the \textit{QAnon}
conspiracy belief.  Similarly, \textit{flat earth} advocates place a large
emphasis on distributing incorrect evidence that is supposed to preemptively
prove the world-view.}.



We see the chance that, should such design ideas become more widespread, their
positive effects transcend their respective interface. Having formed a correct
mental model of one system and being made aware of non-expertise in certain
areas, users are able to carry the same understanding onto other applications.
This is especially important in cases where negative implications are
deliberately hidden.

Finally, we want to stress that maintaining a low bar of entry for
everything digital always needs to be pursued. Our ideas merely serve to prevent a
stagnation concerning the level of insight people can have into their systems.
Going further, measures towards preserving universal ease of use when employing
such design approaches are an important cornerstone. Interfaces themselves must
not become more difficult to understand or use, regardless of any factors that
users may be affected by. In the same vein, the usefulness of any auxiliary
information that is presented must not be lost on users.

\section{Future Research}


The realm of users that are neither novices nor experts remains largely underexplored in
current HCI research. Acquisition of deeper insight
is often externalized to educational facilities or only made possible by
actively searching out information. Integrating a presentation of the presence
of lower levels of abstraction directly inside interface designs is a novel
concept for which we assume a large body of research needs to be conducted.


To conclude, we list a number of research questions that we deem worth exploring as
an initial starting point: \begin{itemize}

    \item Can moments be inferred in which both the application context and a
        user's mental state allow teaching short bits of information?

    \item To what extent can the functionality of lower layers be introduced
        via interface design?

    \item Are users incentivised to actively seek out information about the
        systems they use, if they are given a digestible subset of
        information beyond their skill set?

    \item What mental models do novice users have about their technologies?

    \item Are there ways to extract mental models by analysing computer usage
        and inferring a user's level of knowledge to detect false assumptions?

    \item Do users significantly change their behaviour concerning their
        selection of preferred content when made aware of different filter
        bubbles?

    \item What roadblocks need to be overcome concerning the inclusivity of
        these approaches? Are there relevant cultural differences? To what
        extent do approaches need to be changed to accommodate for disabilities?

\end{itemize}

\bibliographystyle{ACM-Reference-Format}
\bibliography{references}


\begin{thebibliography}{20}


\ifx \showCODEN    \undefined \def \showCODEN     #1{\unskip}     \fi
\ifx \showDOI      \undefined \def \showDOI       #1{#1}\fi
\ifx \showISBNx    \undefined \def \showISBNx     #1{\unskip}     \fi
\ifx \showISBNxiii \undefined \def \showISBNxiii  #1{\unskip}     \fi
\ifx \showISSN     \undefined \def \showISSN      #1{\unskip}     \fi
\ifx \showLCCN     \undefined \def \showLCCN      #1{\unskip}     \fi
\ifx \shownote     \undefined \def \shownote      #1{#1}          \fi
\ifx \showarticletitle \undefined \def \showarticletitle #1{#1}   \fi
\ifx \showURL      \undefined \def \showURL       {\relax}        \fi
\providecommand\bibfield[2]{#2}
\providecommand\bibinfo[2]{#2}
\providecommand\natexlab[1]{#1}
\providecommand\showeprint[2][]{arXiv:#2}

\bibitem[\protect\citeauthoryear{Almuhimedi, Schaub, Sadeh, Adjerid, Acquisti,
  Gluck, Cranor, and Agarwal}{Almuhimedi et~al\mbox{.}}{2015}]%
        {10.1145/2702123.2702210}
\bibfield{author}{\bibinfo{person}{Hazim Almuhimedi}, \bibinfo{person}{Florian
  Schaub}, \bibinfo{person}{Norman Sadeh}, \bibinfo{person}{Idris Adjerid},
  \bibinfo{person}{Alessandro Acquisti}, \bibinfo{person}{Joshua Gluck},
  \bibinfo{person}{Lorrie~Faith Cranor}, {and} \bibinfo{person}{Yuvraj
  Agarwal}.} \bibinfo{year}{2015}\natexlab{}.
\newblock \showarticletitle{Your Location Has Been Shared 5,398 Times! A Field
  Study on Mobile App Privacy Nudging}. In
  \bibinfo{booktitle}{\emph{Proceedings of the 33rd Annual ACM Conference on
  Human Factors in Computing Systems}} (Seoul, Republic of Korea)
  \emph{(\bibinfo{series}{CHI '15})}. \bibinfo{publisher}{Association for
  Computing Machinery}, \bibinfo{address}{New York, NY, USA},
  \bibinfo{pages}{787–796}.
\newblock
\showISBNx{9781450331456}
\urldef\tempurl%
\url{https://doi.org/10.1145/2702123.2702210}
\showDOI{\tempurl}


\bibitem[\protect\citeauthoryear{Bennett, Maton, and Kervin}{Bennett
  et~al\mbox{.}}{2008}]%
        {2008digitalnativedebate}
\bibfield{author}{\bibinfo{person}{Sue Bennett}, \bibinfo{person}{Karl Maton},
  {and} \bibinfo{person}{Lisa Kervin}.} \bibinfo{year}{2008}\natexlab{}.
\newblock \showarticletitle{The ‘digital natives’ debate: A critical review
  of the evidence}.
\newblock \bibinfo{journal}{\emph{British Journal of Educational Technology}}
  \bibinfo{volume}{39}, \bibinfo{number}{5} (\bibinfo{year}{2008}),
  \bibinfo{pages}{775--786}.
\newblock
\urldef\tempurl%
\url{https://doi.org/10.1111/j.1467-8535.2007.00793.x}
\showDOI{\tempurl}


\bibitem[\protect\citeauthoryear{Bucher}{Bucher}{2017}]%
        {algorithmicimaginary}
\bibfield{author}{\bibinfo{person}{Taina Bucher}.}
  \bibinfo{year}{2017}\natexlab{}.
\newblock \showarticletitle{The algorithmic imaginary: exploring the ordinary
  affects of Facebook algorithms}.
\newblock \bibinfo{journal}{\emph{Information, Communication \& Society}}
  \bibinfo{volume}{20}, \bibinfo{number}{1} (\bibinfo{year}{2017}),
  \bibinfo{pages}{30--44}.
\newblock
\urldef\tempurl%
\url{https://doi.org/10.1080/1369118X.2016.1154086}
\showDOI{\tempurl}


\bibitem[\protect\citeauthoryear{Bucher, Helmond, et~al\mbox{.}}{Bucher
  et~al\mbox{.}}{2017}]%
        {bucher2017affordances}
\bibfield{author}{\bibinfo{person}{Taina Bucher}, \bibinfo{person}{Anne
  Helmond}, {et~al\mbox{.}}} \bibinfo{year}{2017}\natexlab{}.
\newblock \showarticletitle{The affordances of social media platforms}.
\newblock \bibinfo{journal}{\emph{The SAGE handbook of social media}}
  (\bibinfo{year}{2017}), \bibinfo{pages}{233--253}.
\newblock


\bibitem[\protect\citeauthoryear{Covington, Adams, and Sargin}{Covington
  et~al\mbox{.}}{2016}]%
        {10.1145/2959100.2959190}
\bibfield{author}{\bibinfo{person}{Paul Covington}, \bibinfo{person}{Jay
  Adams}, {and} \bibinfo{person}{Emre Sargin}.}
  \bibinfo{year}{2016}\natexlab{}.
\newblock \showarticletitle{Deep Neural Networks for YouTube Recommendations}.
  In \bibinfo{booktitle}{\emph{Proceedings of the 10th ACM Conference on
  Recommender Systems}} (Boston, Massachusetts, USA)
  \emph{(\bibinfo{series}{RecSys '16})}. \bibinfo{publisher}{Association for
  Computing Machinery}, \bibinfo{address}{New York, NY, USA},
  \bibinfo{pages}{191–198}.
\newblock
\showISBNx{9781450340359}
\urldef\tempurl%
\url{https://doi.org/10.1145/2959100.2959190}
\showDOI{\tempurl}


\bibitem[\protect\citeauthoryear{Davenport and Beck}{Davenport and
  Beck}{2001}]%
        {attentioneconomy}
\bibfield{author}{\bibinfo{person}{Thomas~H. Davenport} {and}
  \bibinfo{person}{John~C. Beck}.} \bibinfo{year}{2001}\natexlab{}.
\newblock \showarticletitle{The Attention Economy}.
\newblock \bibinfo{journal}{\emph{Ubiquity}} \bibinfo{volume}{2001},
  \bibinfo{number}{May} (\bibinfo{date}{May} \bibinfo{year}{2001}),
  \bibinfo{pages}{1–es}.
\newblock
\urldef\tempurl%
\url{https://doi.org/10.1145/376625.376626}
\showDOI{\tempurl}


\bibitem[\protect\citeauthoryear{Dingler, Funk, and Alt}{Dingler
  et~al\mbox{.}}{2015}]%
        {Dingler2015Proxemics}
\bibfield{author}{\bibinfo{person}{Tilman Dingler}, \bibinfo{person}{Markus
  Funk}, {and} \bibinfo{person}{Florian Alt}.} \bibinfo{year}{2015}\natexlab{}.
\newblock \showarticletitle{Interaction Proxemics: Combining Physical Spaces
  for Seamless Gesture Interaction}. In \bibinfo{booktitle}{\emph{Proceedings
  of the 4th International Symposium on Pervasive Displays}} (Saarbruecken,
  Germany) \emph{(\bibinfo{series}{PerDis '15})}. \bibinfo{publisher}{ACM},
  \bibinfo{address}{New York, NY, USA}, \bibinfo{pages}{107--114}.
\newblock
\showISBNx{978-1-4503-3608-6}
\urldef\tempurl%
\url{https://doi.org/10.1145/2757710.2757722}
\showDOI{\tempurl}


\bibitem[\protect\citeauthoryear{Edwards and Veale}{Edwards and Veale}{2017}]%
        {edwards2017slave}
\bibfield{author}{\bibinfo{person}{Lilian Edwards} {and}
  \bibinfo{person}{Michael Veale}.} \bibinfo{year}{2017}\natexlab{}.
\newblock \showarticletitle{Slave to the algorithm: Why a right to an
  explanation is probably not the remedy you are looking for}.
\newblock \bibinfo{journal}{\emph{Duke L. \& Tech. Rev.}}  \bibinfo{volume}{16}
  (\bibinfo{year}{2017}), \bibinfo{pages}{18}.
\newblock
\urldef\tempurl%
\url{https://doi.org/10.2139/ssrn.2972855}
\showDOI{\tempurl}


\bibitem[\protect\citeauthoryear{Faridani, Bitton, Ryokai, and
  Goldberg}{Faridani et~al\mbox{.}}{2010}]%
        {10.1145/1753326.1753502}
\bibfield{author}{\bibinfo{person}{Siamak Faridani}, \bibinfo{person}{Ephrat
  Bitton}, \bibinfo{person}{Kimiko Ryokai}, {and} \bibinfo{person}{Ken
  Goldberg}.} \bibinfo{year}{2010}\natexlab{}.
\newblock \bibinfo{booktitle}{\emph{Opinion Space: A Scalable Tool for Browsing
  Online Comments}}.
\newblock \bibinfo{publisher}{Association for Computing Machinery},
  \bibinfo{address}{New York, NY, USA}, \bibinfo{pages}{1175–1184}.
\newblock
\showISBNx{9781605589299}
\urldef\tempurl%
\url{https://doi.org/10.1145/1753326.1753502}
\showURL{%
\tempurl}


\bibitem[\protect\citeauthoryear{Gibbs, Moore, Steel, and McKinnon}{Gibbs
  et~al\mbox{.}}{2017}]%
        {GIBBS2017589}
\bibfield{author}{\bibinfo{person}{Shirley Gibbs}, \bibinfo{person}{Kevin
  Moore}, \bibinfo{person}{Gary Steel}, {and} \bibinfo{person}{Alan McKinnon}.}
  \bibinfo{year}{2017}\natexlab{}.
\newblock \showarticletitle{The Dunning-Kruger Effect in a workplace computing
  setting}.
\newblock \bibinfo{journal}{\emph{Computers in Human Behavior}}
  \bibinfo{volume}{72} (\bibinfo{year}{2017}), \bibinfo{pages}{589 -- 595}.
\newblock
\showISSN{0747-5632}
\urldef\tempurl%
\url{https://doi.org/10.1016/j.chb.2016.12.084}
\showDOI{\tempurl}


\bibitem[\protect\citeauthoryear{Hart, Albarracín, Eagly, Brechan, Lindberg,
  and Merrill}{Hart et~al\mbox{.}}{2009}]%
        {10.1037/a0015701}
\bibfield{author}{\bibinfo{person}{William Hart}, \bibinfo{person}{Dolores
  Albarracín}, \bibinfo{person}{Alice~H. Eagly}, \bibinfo{person}{Inge
  Brechan}, \bibinfo{person}{Matthew~J. Lindberg}, {and} \bibinfo{person}{Lisa
  Merrill}.} \bibinfo{year}{2009}\natexlab{}.
\newblock \showarticletitle{Feeling validated versus being correct: A
  meta-analysis of selective exposure to information}.
\newblock \bibinfo{journal}{\emph{Psychological Bulletin}}
  \bibinfo{volume}{135}, \bibinfo{number}{4} (\bibinfo{year}{2009}),
  \bibinfo{pages}{555--588}.
\newblock
\urldef\tempurl%
\url{https://doi.org/10.1037/a0015701}
\showDOI{\tempurl}


\bibitem[\protect\citeauthoryear{Mahmood}{Mahmood}{2016}]%
        {mahmood2016dunningkruger}
\bibfield{author}{\bibinfo{person}{Khalid Mahmood}.}
  \bibinfo{year}{2016}\natexlab{}.
\newblock \showarticletitle{Do People Overestimate Their Information Literacy
  Skills? A Systematic Review of Empirical Evidence on the Dunning-Kruger
  Effect}.
\newblock \bibinfo{journal}{\emph{Communications in Information Literacy}}
  \bibinfo{volume}{10}, \bibinfo{number}{2} (\bibinfo{year}{2016}),
  \bibinfo{pages}{199--213}.
\newblock
\urldef\tempurl%
\url{https://doi.org/10.15760/comminfolit.2016.10.2.24}
\showDOI{\tempurl}


\bibitem[\protect\citeauthoryear{Maier and Seligman}{Maier and
  Seligman}{1976}]%
        {learnedhelplessness}
\bibfield{author}{\bibinfo{person}{Steven~F. Maier} {and}
  \bibinfo{person}{Martin~E. Seligman}.} \bibinfo{year}{1976}\natexlab{}.
\newblock \showarticletitle{Learned helplessness: Theory and evidence}.
\newblock \bibinfo{journal}{\emph{Journal of Experimental Psychology: General}}
  \bibinfo{volume}{105}, \bibinfo{number}{1} (\bibinfo{year}{1976}),
  \bibinfo{pages}{3–46}.
\newblock
\urldef\tempurl%
\url{https://doi.org/10.1037/0096-3445.105.1.3}
\showDOI{\tempurl}


\bibitem[\protect\citeauthoryear{Munson, Lee, and Resnick}{Munson
  et~al\mbox{.}}{2013}]%
        {Munson_Lee_Resnick_2013}
\bibfield{author}{\bibinfo{person}{Sean Munson}, \bibinfo{person}{Stephanie
  Lee}, {and} \bibinfo{person}{Paul Resnick}.} \bibinfo{year}{2013}\natexlab{}.
\newblock \showarticletitle{Encouraging Reading of Diverse Political Viewpoints
  with a Browser Widget}.
\newblock \bibinfo{journal}{\emph{Proceedings of the International AAAI
  Conference on Web and Social Media}} \bibinfo{volume}{7}, \bibinfo{number}{1}
  (\bibinfo{date}{Jun.} \bibinfo{year}{2013}).
\newblock
\urldef\tempurl%
\url{https://ojs.aaai.org/index.php/ICWSM/article/view/14429}
\showURL{%
\tempurl}


\bibitem[\protect\citeauthoryear{Pariser}{Pariser}{2011}]%
        {Pariser2011TheFB}
\bibfield{author}{\bibinfo{person}{Eli Pariser}.}
  \bibinfo{year}{2011}\natexlab{}.
\newblock \bibinfo{booktitle}{\emph{The Filter Bubble: What the Internet Is
  Hiding from You}}.
\newblock \bibinfo{publisher}{Penguin Books}.
\newblock


\bibitem[\protect\citeauthoryear{Rader, Cotter, and Cho}{Rader
  et~al\mbox{.}}{2018}]%
        {10.1145/3173574.3173677}
\bibfield{author}{\bibinfo{person}{Emilee Rader}, \bibinfo{person}{Kelley
  Cotter}, {and} \bibinfo{person}{Janghee Cho}.}
  \bibinfo{year}{2018}\natexlab{}.
\newblock \showarticletitle{Explanations as Mechanisms for Supporting
  Algorithmic Transparency}. In \bibinfo{booktitle}{\emph{Proceedings of the
  2018 CHI Conference on Human Factors in Computing Systems}} (Montreal QC,
  Canada) \emph{(\bibinfo{series}{CHI '18})}. \bibinfo{publisher}{Association
  for Computing Machinery}, \bibinfo{address}{New York, NY, USA},
  \bibinfo{pages}{1–13}.
\newblock
\showISBNx{9781450356206}
\urldef\tempurl%
\url{https://doi.org/10.1145/3173574.3173677}
\showDOI{\tempurl}


\bibitem[\protect\citeauthoryear{Stroud}{Stroud}{2010}]%
        {10.1111/j.1460-2466.2010.01497.x}
\bibfield{author}{\bibinfo{person}{Natalie~Jomini Stroud}.}
  \bibinfo{year}{2010}\natexlab{}.
\newblock \showarticletitle{{Polarization and Partisan Selective Exposure}}.
\newblock \bibinfo{journal}{\emph{Journal of Communication}}
  \bibinfo{volume}{60}, \bibinfo{number}{3} (\bibinfo{date}{08}
  \bibinfo{year}{2010}), \bibinfo{pages}{556--576}.
\newblock
\showISSN{0021-9916}
\urldef\tempurl%
\url{https://doi.org/10.1111/j.1460-2466.2010.01497.x}
\showDOI{\tempurl}


\bibitem[\protect\citeauthoryear{Thompson and Warzel}{Thompson and
  Warzel}{2021}]%
        {nytinsurrection}
\bibfield{author}{\bibinfo{person}{Stuart~A. Thompson} {and}
  \bibinfo{person}{Charlie Warzel}.} \bibinfo{year}{2021}\natexlab{}.
\newblock \bibinfo{booktitle}{\emph{Opinion | They Used to Post Selfies. Now
  They’re Trying to Reverse the Election.}}
\newblock
\urldef\tempurl%
\url{https://www.nytimes.com/2021/01/14/opinion/facebook-far-right.html}
\showURL{%
\tempurl}
\newblock
\shownote{(Retrieved 02/05/2021).}


\bibitem[\protect\citeauthoryear{Union}{Union}{2018}]%
        {pub_series-dataset-64cb0e71-en}
\bibfield{author}{\bibinfo{person}{International~Telecommunication Union}.}
  \bibinfo{year}{2018}\natexlab{}.
\newblock \showarticletitle{ICT Indicators (Edition 2018/1)}.
\newblock  (\bibinfo{year}{2018}).
\newblock
\urldef\tempurl%
\url{https://doi.org/11.1002/pub_series/dataset/64cb0e71-en}
\showDOI{\tempurl}


\bibitem[\protect\citeauthoryear{Yardi and Boyd}{Yardi and Boyd}{2010}]%
        {10.1177/0270467610380011}
\bibfield{author}{\bibinfo{person}{Sarita Yardi} {and} \bibinfo{person}{Danah
  Boyd}.} \bibinfo{year}{2010}\natexlab{}.
\newblock \showarticletitle{Dynamic Debates: An Analysis of Group Polarization
  Over Time on Twitter}.
\newblock \bibinfo{journal}{\emph{Bulletin of Science, Technology \& Society}}
  \bibinfo{volume}{30}, \bibinfo{number}{5} (\bibinfo{year}{2010}),
  \bibinfo{pages}{316--327}.
\newblock
\urldef\tempurl%
\url{https://doi.org/10.1177/0270467610380011}
\showDOI{\tempurl}


\end{thebibliography}

\end{document}